\pdfoutput=1 

\documentclass[aps,a4paper,superscriptaddress,twocolumn,prb,showkeys,showpacs,balancelastpage]{revtex4}

\usepackage[pdftex]{graphicx}
\usepackage{hyperref}
\usepackage{natbib}
\usepackage{amsmath}
\hypersetup{colorlinks=true,linkcolor=blue,citecolor=blue,filecolor=blue,urlcolor=blue}

\begin{document}
  \setcounter{page}{1}
  \date{\today}

\title{Zero-bias conductance peak and Josephson effect in graphene-NbTiN junctions}

\author{M. Popinciuc}
\altaffiliation{Contributed equally to this work}
\affiliation{Kavli Institute of Nanoscience, TU Delft, 2600 GA Delft, The Netherlands}

\author{V. E. Calado}
\altaffiliation{Contributed equally to this work}
\affiliation{Kavli Institute of Nanoscience, TU Delft, 2600 GA Delft, The Netherlands}

\author{X.L. Liu}
\affiliation{Kavli Institute of Nanoscience, TU Delft, 2600 GA Delft, The Netherlands}

\author{A. R. Akhmerov}
\affiliation{Instituut-Lorentz, Universiteit Leiden, 2300 RA Leiden, The Netherlands}

\author{T. M. Klapwijk}
\affiliation{Kavli Institute of Nanoscience, TU Delft, 2600 GA Delft, The Netherlands}

\author{L. M. K. Vandersypen}
\email[]{L.M.K.Vandersypen@tudelft.nl}
\affiliation{Kavli Institute of Nanoscience, TU Delft, 2600 GA Delft, The Netherlands}

\begin{abstract}
We report electronic transport measurements of graphene contacted by NbTiN electrodes, which at low temperature remain superconducting up to at least 11 Tesla. In devices with a single superconducting contact, we find a more than twofold enhancement of the conductance at zero bias, which we interpret in terms of reflectionless tunneling. In devices with two superconducting contacts, we observe the Josephson effect, bipolar supercurrents and Fraunhofer patterns.

\end{abstract}


\maketitle


\section{Introduction}
Graphene has attracted attention since its isolation from graphite\cite{Novoselov2004} and much of its electronic properties are now well understood \cite{CastroNetoRMP2009}.
Graphene differs from conventional two-dimensional electron systems by its linear dispersion relation at low energies and a vanishing band gap.
There has been a lot of interest in studying quantum transport in graphene (G) contacted by superconductors (S).
So far, most experiments have focused on electrical transport in SGS Josephson junctions.
Tunable Josephson supercurrents\cite{Heersche2007}, multiple Andreev reflections\cite{Miao2007,Shailos2007,Du2008}, Andreev bounds states\cite{Dirks2011}, phase diffusion phenomena\cite{Borzenets2011}, macroscopic quantum tunnelling\cite{Lee2011}, and superconducting phase transitions \cite{Kessler2010,Allain2011} have been observed in devices employing Al\cite{Heersche2007,Miao2007,Du2008,Miao2009,Girit2009,Chialvo2010,Trbovic2010,Kanda2010,Choi2010}, W\cite{Shailos2007}, Ta\cite{OAristizabal2009}, Pb\cite{Borzenets2011,Coskun2011,Dirks2011}, PbIn\cite{Lee2011}, Sn\cite{Kessler2010,Allain2011} as superconductors.
An interesting and experimentally yet unexplored direction is to demonstrate specular Andreev reflection \cite{Beenakker2006} by realizing a superconducting gap larger than the potential fluctuations in the graphene. Another interesting possibility is to study the interplay between superconductivity and the quantum Hall effect \cite{Ma1993,Chtchelkatchev2007,Khaymovich2010,Stone2011,Ostaay2011} which requires contacting high-mobility graphene with a superconductor with a large critical magnetic field.  While we were finalizing the manuscript, measurements on graphene with Nb and ReW contacts at high magnetic field were posted, see Ref. \cite{Komatsu2012}.

In this study we report electronic transport through NbTiN based SGN and SGS junctions, where N is a normal, non-superconducting metal electrode.
NbTiN has a large gap, a high critical temperature ($T_C$) and a high upper critical perpendicular magnetic field ($B_{C_2}^\perp$). We present electrical measurements in a field-effect geometry as a function of source-drain bias, temperature ($50$~mK-$5$~K) and external magnetic field ($0$-$11$~T) applied perpendicular to the sample. 
In SGN devices at sub-Kelvin temperatures and up to moderate magnetic fields, we observe a zero-bias conductance peak. We analyze this peak in terms of reflectionless tunneling.\cite{vanWees1992,Marmorkos1993,Giazotto2001} In SGS devices, we observe gate-tunable supercurrents and discuss their magnetic field response, which exhibits characteristic Fraunhofer patterns. 

The remainder of the manuscript is organized as follows. In Sec.~\ref{sec:fab} we describe the details of the sample fabrication. We discuss the measurements in Sec.~\ref{sec:measurements}, and we conclude in Sec.~\ref{sec:conclusion}.

\section{Sample fabrication}
\label{sec:fab}
Graphene samples were prepared on Si substrates with a $285$~nm thick oxide layer by mechanical exfoliation from natural graphite (NGS NaturGraphit GmbH).
Graphene monolayers were selected by optical contrast.\cite{Blake2007}
The electrodes were defined by standard electron beam lithography.
The normal contacts consist of a $8$~nm Ti adhesion layer with a $50$~nm AuPd alloy on top.
The $\sim30$~nm thick NbTiN superconducting contacts were made by DC sputtering of a NbTi target ($30\%$ Ti, $70\%$ Nb atomic percentage) in a Ar/N$_2$ plasma in a Nordiko-2000 system using an unbalanced parallel plate configuration.\cite{Iosad1999}
The Ar and N2 flows were $100$~sccm and $4$~sccm and the deposition pressure was about $6$~mTorr. The deposition conditions were optimized for producing high quality NbTiN films with low stress\cite{Iosad2000,IosadThesis} and a $T_C$ of about $13$~K corresponding to a BCS superconducting gap of $2$~meV.
The electrical measurements indicate an upper critical perpendicular magnetic field $B_{C_2}^\perp$ in excess of $11$~T at $50$~mK (the exact value is not known).

We investigated three ways of contacting graphene with NbTiN: (1) direct sputtering of NbTiN on the graphene, (2) sputtering of NbTiN on a Ti protective layer and (3) sputtering of NbTiN on a Ti/Au protective layer.
Direct sputtering of NbTiN on graphene leads to very high contact resistances of hundreds of k$\Omega$ (measurements of these devices are not discussed further).
This is attributed to damage to the graphene layer underneath the S electrode due to its exposure to the sputtering plasma and/or the highly energetic particles involved in the sputtering process (DC voltages are of the order of 380~V).
In order to prevent this problem we fabricated devices in which we protected the graphene by two different approaches.
For the type A (Fig.~\ref{RT:fig:1}a) devices, we first covered graphene with a thin layer of Ti ($10$~nm) evaporated in an Eva 450 e-beam evaporator, which involves particles with energies of the order of only 1~eV.
Next, in 3 to 5 minutes the sample was transferred through air into the sputtering system, which unavoidably leads to oxidation of the Ti.
Prior to the sputter deposition of NbTiN, the oxidized Ti was cleaned by an Ar RF plasma of $200$~W for $5$~min in a pressure of $6$~mTorr.
This cleaning procedure leaves about $7$~nm Ti.
For the type B devices (Fig.~\ref{RT:fig:6}a) we e-beam evaporated Ti($2$~nm)/Au($2.5$~nm) as a protective layer. Since no oxide is expected to form during the sample transfer, no RF plasma cleaning was done before depositing NbTiN.

\section{Measurements}
\label{sec:measurements}
The electrical transport measurements were performed in a $^3$He/$^4$He dilution refrigerator (Leiden Cryogenics) with a base temperature of $50$~mK using standard low frequency lock-in technique which allowed simultaneous measurement of the current-voltage characteristic and the differential resistance/conductance.
The wiring between electronics and samples involved pi filters at room temperature and RC and Cu-powder filters at the mixing chamber stage.
All measurements were done in a two-terminal configuration and the data presented here were corrected for the RC filter series resistance of $2.5$~k$\Omega$ per filter.

\begin{figure}[top!] 
	\centering
	\includegraphics{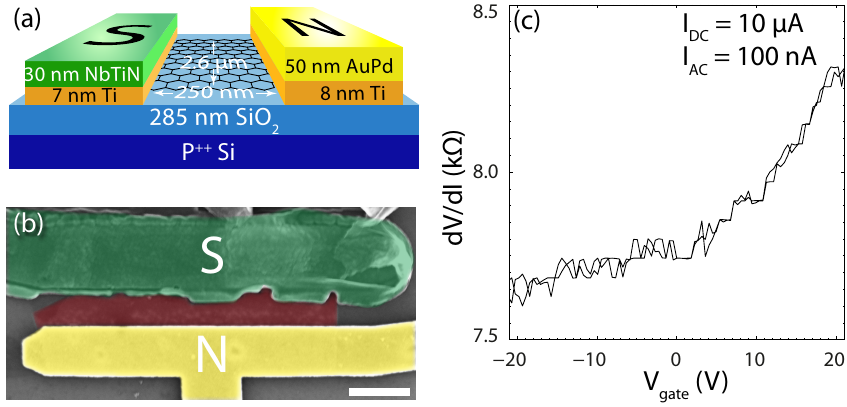}
	\caption{Type A SGN device with 7~nm Ti. (a) Schematic showing dimensions and electrodes configuration. (b) False color SEM image (the white scale bar represents 500~nm).
(c) $dV/dI$ versus $V_{gate}$ at I$_{DC}=10~\mu$A and I$_{AC}=100$~nA in two gate sweep directions.\label{RT:fig:1}}
	
\end{figure}

\begin{figure}[top!]
	\centering
	\includegraphics{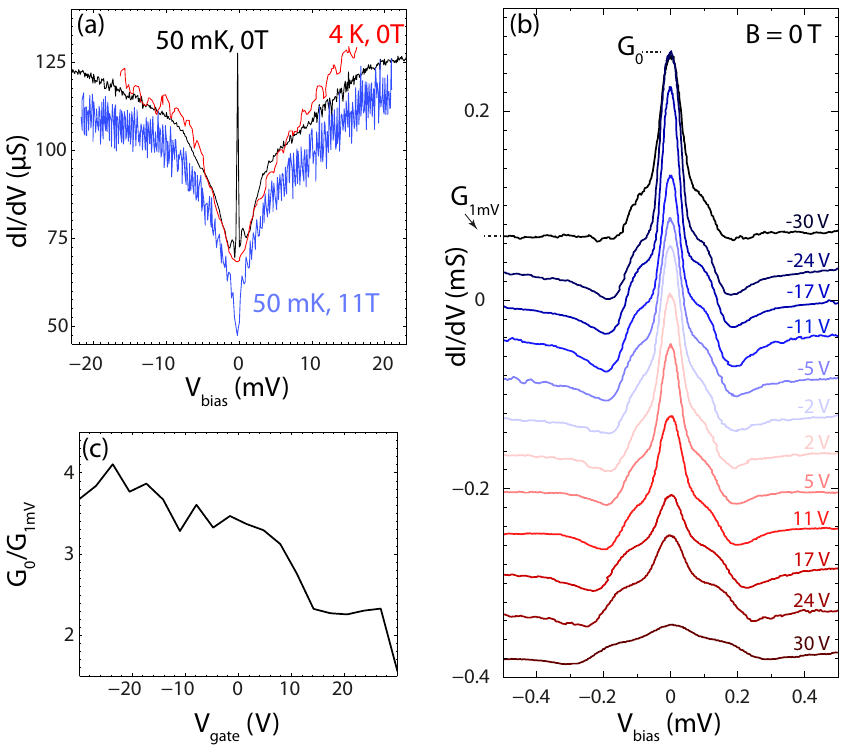}
	\caption{Measurements of the SGN device shown in Fig. 1. (a) $dI/dV$ as a function of $V_{bias}$ for $V_{gate}=0$~V at $4$~K ($B=0$~T) and $50$~mK (for $B=0$~T and $B=11$~T). (b) $dI/dV$ vs. bias voltage at various gate voltages indicated by labels ($T=50$~mK, $B=0$~T). The curves are offset in 40$~\mu$S steps; the top curve has no offset. (c) Zero-bias $dI/dV$ (G$_0$) normalized to the value at 1~mV vs. $V_{gate}$. The zero bias conductance is enhanced by a factor larger than two at almost all gate voltages. \label{RT:fig:2}}
\end{figure}

In Fig.~\ref{RT:fig:1}a and \ref{RT:fig:1}b we show a schematic and a false-color scanning electron micrograph (SEM) of a type A device. The graphene flake is $2.6~\mu$m wide.
The S and N electrodes completely overlap the flake and are separated by about $250$~nm (edge to edge). In Fig.~\ref{RT:fig:2}a we present two-terminal differential resistance ($dV/dI$) measurement at $50$~mK as a function of the back-gate voltage $V_{gate}$ for the as-fabricated SGN junction.
The measurement was done with an AC current of $100$~nA superimposed on a DC current of $10~\mu$A. The $dV/dI$ is rather high over the entire gate voltage range: given the device dimensions, the doping level and an estimated carrier mobility of $\sim2000$~cm$^2$V$^{-1}$s$^{-1}$ observed in a nearby device on the same flake, the graphene differential resistance should be well below 1~k$\Omega$.  Other devices made in the same batch show similar or even higher differential resistances, but type A devices with a thicker Ti protective layer (about 20~nm) and type B devices (without RF plasma cleaning) do not.
Three-terminal measurements of the S and N contact resistances (to the graphene covered by the contacts) show values of 400~$\Omega$ and 700$~\Omega$, respectively. Therefore, we conclude that most of the resistance is due the transition region from the covered graphene to the uncovered graphene (denoted as G$'$). We believe that stress in the NbTiN films and/or RF plasma cleaning may cause damage to the graphene area around the S contacts.

In Fig.\ref{RT:fig:2}a we show the $dI/dV$ for $V_{gate}=0$~V as a function of voltage bias at $60$~mK (for two perpendicular magnetic field values of $0$~T and $11$~T) and at $4$~K for 0~T.
We see that $dI/dV$ is non-constant over the entire bias-voltage range of $\pm 20$~mV. Since the NbTiN gap is only $\sim 2$~mV, the non-constant $dI/dV$ above 2~mV cannot be attributed to the superconducting contacts. Indeed, when normalized to a high bias value of $10-20$~mV, the three dI/dV traces in Fig.\ref{RT:fig:2}a fall on top of each other except for a bias range of less than $2$~mV.
The observed behavior is also incompatible with known mechanisms that lead to a bias-dependent graphene resistance.\cite{Barreiro2009,Vandecasteele2010,Perebeinos2010} Possibly it is caused by the damaged graphene near the contacts.

Fig.~\ref{RT:fig:2}b shows $dI/dV$'s at low bias for several gate voltages.
We observe a zero bias conductance peak for energies $E_C$ smaller than 0.1~meV and a conductance dip at energies $E_D$ of around 0.2~meV. Both features are present at all gate voltages. The zero bias conductance is enhanced by more than a factor of two compared with its value at $\sim 1$~mV for almost all the back gate voltages investigated, see Fig~\ref{RT:fig:2}c. Ignoring possible magnetic moments \cite{Nair2012,Balatsky2006}, to the best of our knowledge only reflectionless tunneling can explain an enhancement a factor of more than two.  Hereby the transmission of a tunnel barrier between a superconductor and a normal metal is enhanced due to the diffusive transport in the normal metal \cite{vanWees1992,Marmorkos1993,Giazotto2001}. 

\begin{figure}[top!]
	\centering
	\includegraphics{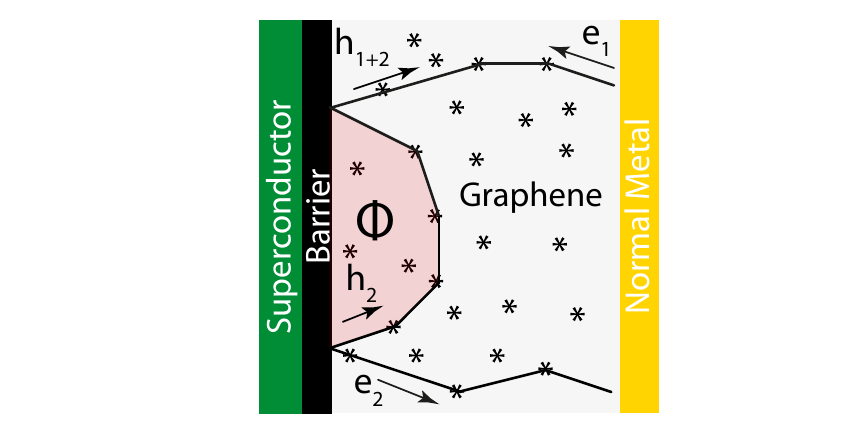}
	\caption{Illustration of reflectionless tunneling.  Quasi-particles move from the right reservoir towards the superconductor through diffusive graphene. A potential barrier exists between graphene and the superconductor. The Andreev reflection probability at an otherwise poorly transparent interface is enhanced due to interference effects (see text for details).	\label{RT:fig:3}}
\end{figure}

Following the semiclassical approach of \citeauthor{vanWees1992} we sketch the principle behind reflectionless tunneling in Fig.~\ref{RT:fig:3}. The quasi-particles move from the right normal metal reservoir towards the S electrode through the diffusive graphene. A potential barrier exists at the graphene/S interface, which in our case is assumed to be due to the damaged graphene around the S electrode.
An electron (e$_1$ in Fig.~\ref{RT:fig:3}) that hits the barrier can be either Andreev reflected as a hole (h$_1$) or normal reflected as an electron, continuing along path 2.
Due to scattering on impurities the electron moving on path 2 has a chance to hit the barrier once again where it can retroreflect as a hole (h$_2$).
Retracing path 2, this second hole reaches the initial point where it can undergo normal reflection.
Constructive interference between the first and the second hole increases the total Andreev reflection probability of the incoming electron e$_1$. In a diffusive sample there are a multitude of such paths and their respective contributions add up leading to an enhancement of the conductance. At finite bias, the phase of h$_2$ averages out and the conductance enhancement disappears.

The order of magnitude of the cutoff energy $E_C$ for which coherence is lost is determined by the phase coherence time $\tau_\phi$ according to $E_C\approx\hbar/\tau_\phi$. Taking $E_C\approx 0.065$~meV from the $dI/dV$ at $V_{gate}=0$~V, we estimate $\tau_\phi\approx 10$~ps which is comparable with the values obtained in Ref.\cite{Tikhonenko2008}. The phase breaking length is given by $L_\phi=\sqrt{D\tau_\phi}$.
Here, $D$ is the diffusion constant given by $D=v_F l/2$, where $l$ is the mean free path and $v_F \approx 10^6$~m~s$^{-1}$ the Fermi velocity.
Taking $\tau_\phi=10$~ps and a mean free path of $\approx 17$~nm (corresponding to a carrier mobility of 1000~cm$^2$/Vs at $30$~V from the Dirac point) gives $L_\phi\approx 280$~nm, which is comparable to the sample length $L=250$~nm.

\begin{figure}[top!]
	\centering
		\includegraphics{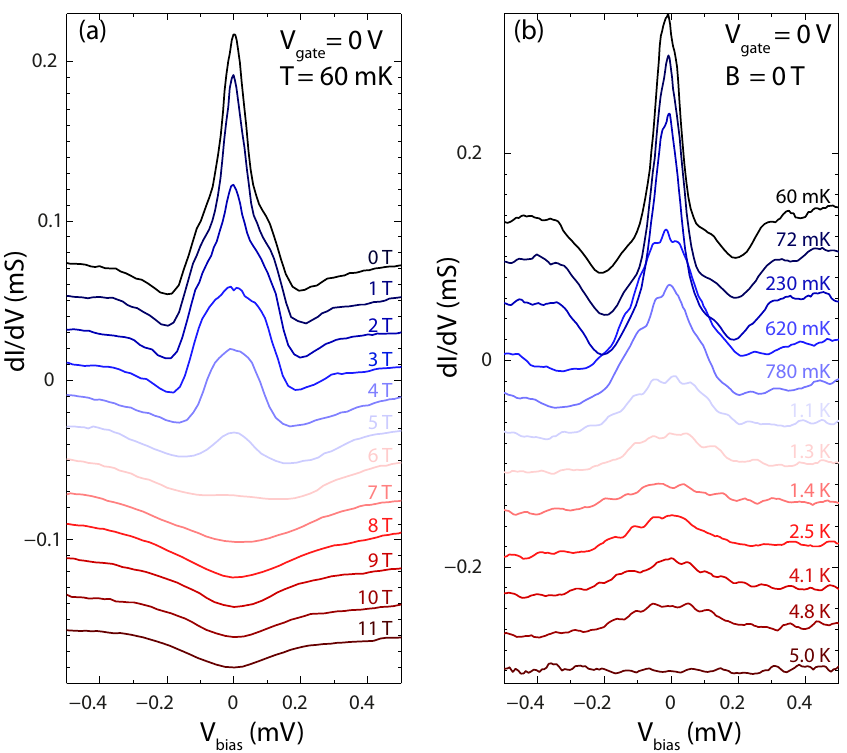}
	
	\caption{ (a) $dI/dV$ vs. bias voltage at $V_{gate}=0$~V for different magnetic fields.  (b) Temperature dependence of the $dI/dV$ at $V_{gate}=0$~V and $B=0$~T. For clarity, the data in both panels is offset vertically by incrementally subtracting a value of $40$~$\mu$S. The topmost curves have no offset. \label{RT:fig:4}}
\end{figure}

Now we turn our attention to the magnetic field and the temperature dependence.
In Fig.~\ref{RT:fig:4}a we show the $dI/dV$'s as a function of bias voltage for various magnetic fields applied perpendicular to the sample surface.
A finite magnetic field breaks the time reversal symmetry and introduces a phase difference between the interfering holes. Coherence is lost when the loop formed by path 2 and the superconductor encloses one flux quantum $\Phi_0=h/e$. Taking $B=4$~T as the field where reflectionless tunneling is suppressed (see Fig.~\ref{RT:fig:4}), we estimate an effective area of $1\times10^3$~nm$^2$. At high magnetic fields one expects Landau levels to develop as the graphene enters the quantum Hall regime.
However since we did not see clear signs of quantum Hall plateaus, it is likely that the disorder in the sample was too high for the quantum Hall effect to develop.

In Fig.~\ref{RT:fig:4}b we plotted the temperature dependence of the $dI/dV$ at $V_{gate}=0$~V and $B=0$~T. The measurements were taken after a mild current annealing step\cite{Moser2007} performed at base temperature.
For about $10$~minutes, we slowly ramped up a DC current applied between the S and N contacts, up to a current density of $4.5$~A~cm$^{-1}$. While this lead to a $\approx 50\%$ increase in the overall conductance, the zero-bias conductance peak and the broader conductance dip remained at about the same energies. This behavior is similar to what was reported in Ref.\cite{OAristizabal2009}. From Fig.~\ref{RT:fig:4}b we observe that the zero-bias peak drops at about 1 K. This thermal energy scale is consistent with $E_c$ extracted from the width of the peak.

\begin{figure}[top!]
	\centering
		\includegraphics{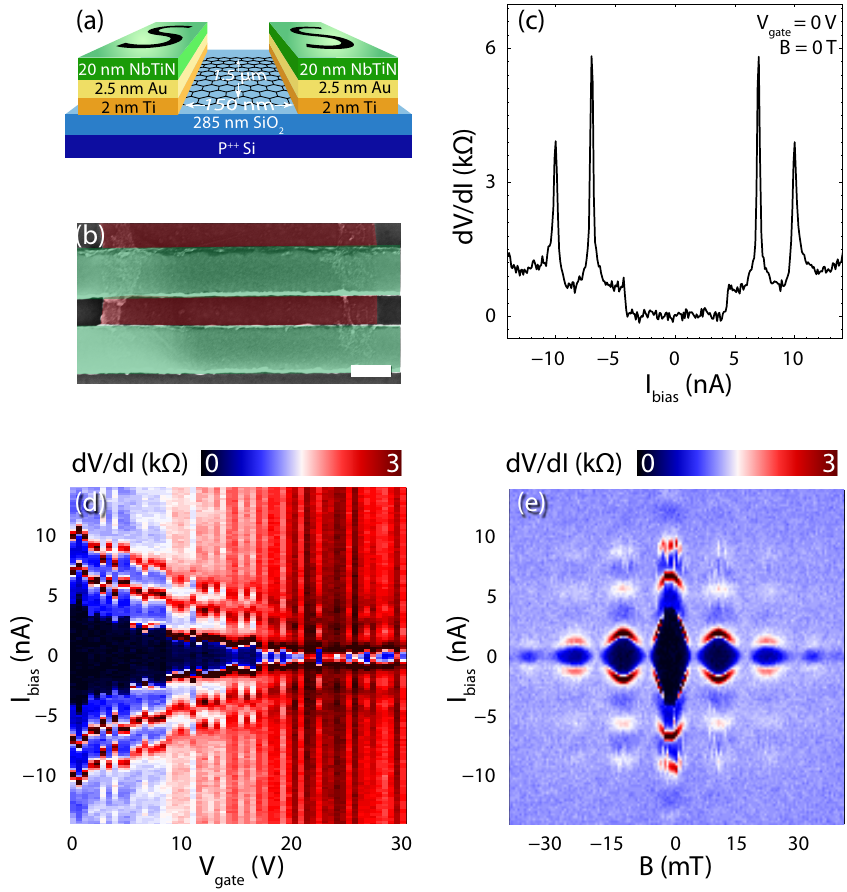}
	\caption{Type B SGS device with Ti/Au/NbTiN electrodes made without RF cleaning prior to depositing NbTiN. (a) Schematic of the device structure with the relevant dimensions. (b) False color SEM image (scale bar $200$~nm). (c) $dV/dI$ at $V_{gate}=0$~V, $B=0$~T. (d) Color plot of $dV/dI$ vs. $I_{bias}$ and $V_{gate}$ at $B=0$~T. (e) Color plot of $dV/dI$ vs. $B$ and $I_{\mbox{bias}}$ at $V_{gate}=0$~V.\label{RT:fig:6}}
\end{figure}

Now, we briefly discuss measurements of SGS junctions.
In a type A SGS device with similar dimensions as the SGN device discussed earlier and made in the same batch, no supercurrent flowing through graphene (Josephson effect) was observed. 
We understand this in terms of the poor transparency of the G-G$'$ interfaces which hinder the diffusion of the Cooper pairs into uncovered graphene.
In another batch, with a thicker Ti layer (about $20$~nm after the RF plasma cleaning), we did observe bipolar supercurrents and Fraunhofer patterns in several graphene junctions for electrode spacings of up to $400$~nm\cite{LiuThesis}. This is indicative of less damage to the G-G$'$ transition region. These data are not discussed further.

In Fig.~\ref{RT:fig:6} we show measurements of a type B device. Given the Ti($2$~nm)/Au($2.5$~nm) protective layer, no oxide is expected to form during the sample transfer through air so we skipped the RF plasma cleaning step before deposition of NbTiN.
Fig.~\ref{RT:fig:6}a and Fig.~\ref{RT:fig:6}b show the schematic and a false color SEM image of the device. The graphene flake is $1.5~\mu$m wide and the electrode separation is $150$~nm (edge to edge).
In Fig.~\ref{RT:fig:6}c we show the $dV/dI$ as a function of bias current at $V_{gate}=0$~V and $B=0$~T.
A critical current $I_C$ of $4$~nA is observed. We note that $I_C$ is relatively small given the junction dimensions and the carrier density. Also, we did not observe supercurrents in type B junctions with $280$~nm or larger electrode separations, possibly due to the weakening of the proximity effect in the Ti/Au bilayer.
The sharp peaks in $dV/dI$ we interpret as self-induced Shapiro steps  since the energies involved closely match those of standing waves formed in the metal box enclosing the sample,  which has a size of $6.0$~cm.\cite{Levinsen1974,Klapwijk1977}
In Fig.~\ref{RT:fig:6}d we show $dV/dI$'s as a function of the gate voltage and bias current as 2D color plot.
The dark region corresponds to a supercurrent through graphene.
The magnitude of the critical current depends on the charge carrier density and decreases when going from metallic conduction towards the charge neutrality point (at $\sim22$~V), in agreement with previous reports\cite{Heersche2007}. In Fig.~\ref{RT:fig:6}e we show $dV/dI$s as a function a perpendicular magnetic field and bias current at zero gate voltage.
The critical current is modulated by the magnetic field revealing the well-known Fraunhofer pattern.
The area extracted from the periodicity of the Fraunhofer pattern amounts to $0.16~\mu$m$^2$ in close agreement with the geometrical area of $0.2~\mu$m$^2$ deduced from the SEM image.
The width of the first lobe is smaller than twice the period of the higher order lobes which indicates non-uniform current flow.

\section{Conclusion}
\label{sec:conclusion}
In this paper we reported electrical transport measurements of SGN and SGS junctions with NbTiN superconducting electrodes.
In SGN devices with a barrier region near the S contacts, we observe a zero-bias conductance peak, which we think may arise from reflectionless tunneling. While we are unable to identify its origin unambiguously, this peak has a striking resemblance to the recently measured zero-bias anomaly in a candidate topological superconductor Cu$_x$Bi$_2$Se$_3$.\cite{Sasaki2011}
In junctions with transparent interfaces, supercurrents were observed for electrode separations of up to $400$~nm.
By improving the fabrication procedure we believe that high carrier mobility graphene samples and clean graphene-NbTiN interfaces can be obtained for experimentally investigating specular Andreev reflection and the interplay between Andreev reflection and the quantum Hall effect.

\begin{acknowledgments}
We acknowledge N. Vercruyssen, D.J. Thoen, T. Zijlstra, A.M. Goossens, A. Barreiro, C.J.H. Keijzers and S.M. Frolov for discussions and technical support. This work is supported by the Dutch Science Foundation NWO/FOM and by the Eurocores program EuroGraphene.
\end{acknowledgments}

\bibliography{Ref}

\end{document}